\documentclass[%
superscriptaddress,
groupedaddress,
reprint,
twocolumn,
 amsmath,amssymb,
 aps,
prb,
]{revtex4}

\usepackage{graphicx}
\usepackage{amsmath}

\begin{document}

\preprint{}

\title{Tin telluride: a weakly co-elastic metal}

\author{E.K.H. Salje}
\affiliation{Department of Earth Sciences, Cambridge University, Downing Street, Cambridge CB2 3EQ, United Kingdom}
\affiliation{Center for Nonlinear Studies,Los Alamos National Laboratory, Los Alamos, NM 87545, USA}

\author{D.J.~Safarik}
\affiliation{Los Alamos National Laboratory, Los Alamos, NM 87545, USA}
\author{K.A.~Modic}
\affiliation{Los Alamos National Laboratory, Los Alamos, NM 87545, USA}
\author{J.E.~Gubernatis}
\affiliation{Los Alamos National Laboratory, Los Alamos, NM 87545, USA}
\author{J.C.~Cooley}
\affiliation{Los Alamos National Laboratory, Los Alamos, NM 87545, USA}
\author{R.D.~Taylor}
\affiliation{Los Alamos National Laboratory, Los Alamos, NM 87545, USA}
\author{B.~Mihaila}
\affiliation{Los Alamos National Laboratory, Los Alamos, NM 87545, USA}
\author{A.~Saxena}
\affiliation{Los Alamos National Laboratory, Los Alamos, NM 87545, USA}
\author{T.~Lookman}
\affiliation{Los Alamos National Laboratory, Los Alamos, NM 87545, USA}
\author{J.L.~Smith}
\affiliation{Los Alamos National Laboratory, Los Alamos, NM 87545, USA}

\author{R.A. Fisher}
\affiliation{Lawrence Berkeley National Laboratory, Berkeley, CA 94720, USA}

\author{M.~Pasternak}
\affiliation{Tel Aviv University, Ramat Aviv 69978, Israel}

\author{C.P.~Opeil}
\affiliation{Department of Physics, Boston College, Chestnut Hill, Massachusetts, USA}

\author{T. Siegrist}
\affiliation{National High Magnetic Field Laboratory, 1800 E Paul Dirac Dr, Tallahassee, Florida, USA}

\author{P.B.~Littlewood}
\affiliation{Cavendish Laboratory, Cambridge University, Madingley Road, Cambridge CB3 0HE, United Kingdom }

\author{J.C.~Lashley}
\affiliation{Los Alamos National Laboratory, Los Alamos, NM 87545, USA}

\date{\today}

\begin{abstract}
We report resonant ultrasound spectroscopy (RUS), dilatometry/magnetostriction, magnetotransport, magnetization, specific-heat, and $^{119}$Sn M\"ossbauer spectroscopy measurements on SnTe and Sn$_{0.995}$Cr$_{0.005}$Te. Hall measurements at $T=77$~K indicate that our Bridgman-grown single crystals have a $p$-type carrier concentration of $3.4 \times 10^{19}$ cm$^{-3}$ and that our Cr-doped crystals have an $n$-type concentration of $5.8 \times 10^{22}$ cm$^{-3}$.  Although our SnTe crystals are diamagnetic over the temperature range $2\, \text{K} \leq T \leq 1100\, \text{K}$, the Cr-doped crystals are room temperature ferromagnets with a Curie temperature of 294~K. For each sample type,  three-terminal capacitive dilatometry measurements detect a subtle 0.5~micron distortion at $T_c \approx 85$~K. Whereas our RUS measurements on SnTe show elastic hardening near the structural transition, pointing to co-elastic behavior, similar measurements on Sn$_{0.995}$Cr$_{0.005}$Te  show a pronounced softening, pointing to ferroelastic behavior.
Effective Debye temperature, $\theta_D$, values of SnTe obtained from $^{119}$Sn M\"ossbauer studies show a hardening of phonons in the range 60--115K ($\theta_D$ = 162K) as compared with the 100--300K range ($\theta_D$ = 150K).
In addition,  a precursor softening extending over approximately 100 K anticipates this collapse at the critical temperature, and quantitative analysis over three decades of its reduced modulus finds $\Delta C_{44}/C_{44}=A|(T-T_0)/T_0|^{-\kappa}$ with $\kappa = 0.50 \pm 0.02 $, a value indicating a three-dimensional softening of phonon branches at a temperature $T_0 \sim 75$~K, considerably below $T_c$. We suggest that the differences in these two types of elastic behaviors lie in the absence of elastic domain wall motion in the one case and their nucleation in the other.
\end{abstract}

\pacs{Valid PACS appear here}
\maketitle

\section{Introduction}

For around 50 years, the IV-VI compounds have provided a basis for semimetals and small-gap semiconductors with high-temperature phases in various crystal structures transforming into other structural phases on lowering the temperature~\cite{volkov82,nimtz83,jantsch83,holder83,pblreview83,Littlewood80,kristoffel88}. The low-temperature phases themselves have often been objects of considerable interest. On structurally transforming, SnTe and GeTe, for example, become polar, with some authors invoking ferroelectricity. In practice the physical properties of members of this family of compounds depend on the value of their excess carrier concentration~\cite{nimtz83,jantsch83,holder83,pblreview83}, although an experimental link to defects is also possible. In fact, several ternary compounds, such as Pb$_{1-x}$Sn$_x$Te and Ge$_{1-x}$Sn$_x$Te, have also received considerable study because changing the carrier concentration was seen as way not only to tailor the size of the band gaps but also to switch the parity of the bands across the gap. A recent interest revival\cite{wuttig,thermoelectrics,spinhall} in these materials is occurring in part because of a perceived opportunity to create novel tunable multifunctional devices on the nano-scale. Just as recently, we started a study of SnTe to exploit another property of this and several other IV-VI compounds; namely, upon doping with a magnetic atom, such as Cr or Mn, the resulting compound becomes ferromagnetic~\cite{ferromagnetic1,ferromagnetic2}. Thus, magnetically-doped SnTe has the potential of being multiferroic with three features -- ferroelasticity, ferroelectricity, and ferromagnetism -- in a considerably simpler crystal structure than the perovskite-structured manganite multiferroics that typically are studied for just two features -- ferroelectricity and ferromagnetism.  It was envisaged that the combination between semi-conducting and ferroic properties is  desirable to investigate the possibility of a new class of ferroic materials.

In this paper, we report the presence of \emph{ferroelasticity} in Cr-doped SnTe below the phase transition point near 100K. Ferroelasticity implies that a spontaneous strain arises below a critical temperature and that the switching of the variants must be allowed. In contrast, SnTe is \emph{co-elastic}, i.e. the appearance of spontaneous strain occurs without variant switching, where the elastic strain energy is insufficient to form ferroelastic domains.

More specifically, the high-temperature phase of SnTe transforms from cubic (rocksalt) to rhombohedral around 100~K, the precise value depending on the excess carrier concentration.  Electronic structure theory points to unsaturated bonding dominated by the $p$-orbitals of Sn and Te with a stoichiometric compound being insulating with an approximate 2~eV gap across the fcc Brillouin zone except around the L-point where the gap is about 0.3~eV~\cite{volkov82,holder83,bandtheory,ARPES}. Across the L-point, the band states have opposite parity and couple to a transverse optical phonon in the [111] direction, a Raman active mode that drives a second-order displacive transition in this direction. The measured size of the displacement is approximately 50 pm. Accompanying it is a shear strain of approximately 0.122$^{\circ}$ that decreases the rhombohedral angle.

\textit{Prima facie}, the transition in SnTe seems to fit the scenario discussed by Anderson and Blount~\cite{anderson65}, whereby a second-order transition with strain must be accompanied by some other internal symmetry change. One possible change is the spontaneous development of a symmetry represented by a polar (vector) order parameter.  In the present case, this polar order parameter is the static displacement associated with the optical phonon. Such a polar order parameter would produce ferroelectricity if there were a spontaneous electrical polarization that can be reversed by the application of an electric field.  The best evidence for electrical polarization in SnTe is its anomalous transverse effective charge inferred from Raman scattering~\cite{pblreview83}. For the observed sizes of the displacement, the effective charge implied by the measured dipole moments is around 4 to 8 times the unit charge available from fully ionized Sn and Te. Another cited feature is the large static and optical frequency dielectric constants of approximately 1000 and 50, respectively, near the structural transition.

In this paper we report a new suite of measurements on SnTe that produced unexpected results. In practice SnTe is always found to have a substantial carrier density, probably from defect-induced carriers, and we confirmed this in our samples. In consequence an internal electrical field cannot be maintained and ferroelectricity as a macroscopic phenomenon is not possible. We did observe weak signatures of a transition, presumably structural, around 100~K; however, instead of ferroelasticity we found co-elasticity. Across the transition, ferroelastic and coelastic crystals exhibit only small changes in physical properties, such as elastic stiffness and thermal expansion, whereas ferroelastics typically exhibit large changes. The co-elastic changes we found for SnTe are likely among the smallest yet observed.

We also performed measurements on Cr-doped SnTe and found several unexpected results. Specifically, we found a 0.5 \% Cr doping produced a room temperature ferromagnet emerging from diamagnetism in the high-temperature phase. More surprisingly, the dilute level of doping changed the material from co-elastic to ferroelastic.

The purpose of this paper is to report these findings. We remark that our samples, as other samples of SnTe reported in the literature, are off-stoichiometry, and hence metallic.  Elsewhere, \cite{ARPES} we have reported angular-resolved photoemission (ARPES) measurements that show over a wide range of temperatures a Fermi surface whose topology changes markedly with electron binding energy. A hole doping of about  $5 \times 10^{19} cm^{-3}$ shifts the fermi energy into the uppermost valence band by several tenths of an eV, consistent with our magnetotransport results. In this paper we focus on other measurements that support the presence of the transition, the metallic behavior, and the weak elastic and thermal anomalies. We remark that the strong carrier-concentration dependence of the structural transition in SnTe is an effect normally not seen in other structurally transforming materials, such as shape-memory alloys where the martensitic phase transition is structurally identical to a ferroelastic transition but where the equivalent Landau free energy change is very large~\cite{martensites}.

It is convenient to begin from a theoretical perspective because we will find it straightforward to interpret some of the results using a Ginzburg-Landau theory. In Section~II, we outline a basic dual order-parameter Landau free-energy functional for studying the phase transition. In Section~III, we discuss how we grew our SnTe crystals and how we performed our resonant ultrasound spectroscopy (RUS), dilatometery, magnetostriction, magnetotransport, specific heat, and $^{119}$Sn M\"ossbauer spectroscopy measurements.  In Section~IV, we report the results of our measurements.  Finally, in Section~V, we give summary remarks.

\section{Theory}

We will discuss the structural phase transitions in pure SnTe and Cr-doped SnTe in terms of a Landau-Ginzburg free energy~\cite{kittel,toledano,Saljebook}.  As standard, we construct this energy by starting from the symmetry of the two phases. The high temperature phase has the rock salt structure (space group Fm3m, site symmetry m3m, Wykoff positions 000 and $(\frac{1}{2} \frac{1}{2}\frac{1}{2}$)),  and the low temperature phase has the rhombohedral structure (space group R3m, site symmetry m-3m, Wykoff positions ($xxx$) with $x=0.004$ and $\frac{1}{2}- 0.004$, rhombohedral angle 59.878 at  zero K)~\cite{Rabe}. By group theory arguments, the phase transition occurs by a two step symmetry reduction~\cite{Bilbao}. The first is the reduction of the cubic phase symmetry to that of a rhombohedral phase (R-3m) via an elastic strain along the cube diagonal. The irreducible representation of this elastic order parameter is $T_{2g}$~\cite{Saljebook}, and this resulting phase has an inversion center. The allowed strain components are $xy$, $xz$, and $yz$. The second symmetry reduction removes the inversion center by a relative displacement of the two sub-lattices whose irreducible representation is $T_{1u}$~\cite{Saljebook}. The basis functions of $T_{1u}$ are $x$, $y$, and $z$.

The primary order parameter describes the polar displacement between the two sublattices of approximately 50~pm. (In SnTe, this represents a one part in 10$^9$ effect.)  This displacement vector is parallel to $\langle111\rangle$. The secondary order parameter is the strain $|e_{xy}|=|e_{xz}|=|e_{yz}|$, leading to an elongation of the unit cell along any of the polar axes. This shear corresponds to an rhombohedral shear angle  of $0.122^{o}$. The conjugated elastic modulus is $C_{44}$, which is expected to be explicitly temperature dependent in the vicinity of the transition, whereas all other elastic moduli remain relatively unaffected by the phase transition~\cite{Saljebook}. Both order parameters act at the origin of the Brillouin zone and maintain the topology of the unit cell.

The Landau approach to this transformation leads to a mean-field approximation where all secondary interactions are treated as perturbations of the primary order parameter~\cite{toledano,Saljebook}.  Band structure calculations\cite{Rabe} report a non--convex energy well that is deep along the polar deformation but is shallow along the spontaneous strain, a situation typical for multiferroic materials where electrical and elastic degrees of freedom mix~\cite{order parameter coupling}. We will follow the treatment of ferroelectric materials and treat the polar order parameter as the driving (primary) order parameter and add the elastic relaxation in the appropriate symmetry-adapted perturbation.  The Gibbs free energy of the zone center cubic-to-rhombohedral transition is
\begin{align}\label{gibbsenergy}
G(P_i,e_{jk})= L( P_i ) & + f_\mathrm{el} (e_{jk} )+ f_\mathrm{coupling}( P_i, e_{jk})
\notag \\
& + f_\mathrm{Ginzburg}(\nabla P_{i}) \>,
\end{align}
where the first term on the right is the non-convex Landau energy, the second term is the elastic energy, the third term is the coupling energy between the polar order parameter $P_i$ and spontaneous strain e$_{jk}$ components, and the last term is the Ginzburg energy which is relevant for the assessment of the domain boundaries. In the following paragraphs, we state expressions for only the first three of these contributions, thereby choosing to describe the physics in a single domain.

The general form of the Landau free energy $L(P_{i})$ for the rhombohedral (m3m) symmetry is\cite{m3m,toledano,Saljebook}
\begin{align}\label{landauenergy}
& L(P) =  \frac{1}{2} \alpha(P_{1}^{2}+P_{2}^{2}+P_{3}^{2}) + \frac{1}{2} B'(P_{1}^{4}+P_{2}^{4}+P_{3}^{4})
\notag \\ &
+ \frac{1}{2} B''(P_{1}^{2} P_{2}^{2}+P_{2}^{2} P_{3}^{2}+ P_{1}^{2} P_{2}^{2}) + \frac{1}{6} C'(P_1P_2P_3)^2
\notag \\ &
+ \frac{1}{6} C''(P_{1}^{2}+P_{2}^{2}+P_{3}^{2})^3
\notag \\ &
+ \frac{1}{6} C'''(P_{1}^{2}+P_{2}^{2}+P_{3}^{2})(P_{1}^{4}+P_{2}^{4}+P_{3}^{4}) \>.
\end{align}
Within one domain, for example, the $(111)$ domain, one component of $P_i$ is non-zero while all others are zero. Thus the mono-domain Landau free energy has the reduced form
\begin{align}\label{monodomainlandau}
 L(P) = \frac{1}{2}\alpha P^2+\frac{1}{4}B P^4 +\frac{1}{6}C P^6 \>,
 \end{align}
with coefficients $B=B(B',B'')$ and $C=C(C',C'',C''')$. This is the simplest Landau free energy for a continuous phase transition when $B > 0$ or $B = 0$  and $C > 0$. The case $B=0$ is called \textit{tricritical} as it  separates a second order ($B > 0$) and first order ($B < 0$) transformation.

The standard assumption in a Landau theory of phase transitions is $\alpha=A|T-T_c|$.
Very close to $T_c$, this assumption imposes a mean-field critical exponent on the thermodynamics, which is generally incorrect. Very far from $T_c$, as $T \to 0$, the temperature dependences of thermodynamic quantities become inconsistent with the requirement that the entropy must approach zero. In the interpretation of our results, we will assume that the structural transition is driven by a single anharmonic soft mode. With this single-mode assumption,
the low-temperature form
\begin{equation}\label{softmodealpha}
\alpha= A\Theta_s [\coth (\Theta_s/T) -\coth (\Theta_s/T_c)] \>,
\end{equation}
exactly follows from a self-consistent phonon approximation to this mode~\cite{Saljebook,dove92}. In various contexts, this form has provided informative fits to low-temperature properties derived from a Landau free energy.

In Eq.~\eqref{softmodealpha}, the parameter $\Theta_s$ is called the saturation temperature. SnTe has a Raman active soft mode in the rhombohedral phase~\cite{softmodes}.  In this case, the saturation temperature is an estimate of its frequency. With these soft optical-mode frequencies~\cite{softmodes}.  We estimate the saturation temperature of SnTe to be between 40~K and 60~K, which is consistent with the value obtained from our thermal expansion data (see Fig.~\ref{excessstrain}).  The phononic saturation temperature based on a Debye temperature would be appropriate for a proper ferroelastic system.

For the $[111]$ domain the spontaneous strain is symmetry constrained with $e_{4} = e_{xy}+e_{xz}+e_{yz}$. The elastic energy is for a single domain crystal $f_\text{el} (e_{4})=\frac{1}{2} C_{44} e_{4}^{2}$, and the coupling energy is $f_\text{coupling}( P_i, e_{ij})= \frac{1}{2} P^{2} e_{4}$.  If we were to relax the free energy with respect to this strain, that is,  solve $dG/de_{4}=0$ and compress \eqref{landauenergy}, the resulting expression would have the same functional form as Eq.~\ref{monodomainlandau}, but with $B$ replaced by a renormalized parameter $B^*$~\cite{Carpenter}. Thus, as typical when a secondary order parameter is present, the nature of the transition remains described by the Landau free-energy form of the primary order parameter but with a shifted critical temperature~\cite{toledano}.

The corollary to the effect of strain coupling in shifting the transition, is that a spontaneous strain appears in response to the order parameter. Because this is a macroscopic effect on the crystal, it is often much easier to observe than the order parameter itself.

\section{Experimental}

We prepared large single crystals (1$\times$1$\times$2~cm$^{3}$) of SnTe  either by fusing the elements in a Bridgman furnace (temperature gradient).
We took advantage of the high vapor pressure of Te and lowered the growth temperature in the Bridgman to 850$^\circ$C.  Powder diffraction and back-reflection Laue images showed that the crystals grew along the $\langle321\rangle$ crystallographic directions, and rocking curves measured by neutron diffraction indicated a small mosaic of 0.5$^\circ$.

We  determined the elastic constants of disk-shaped SnTe  single crystals with resonant ultrasound spectroscopy (RUS)~\cite{migliori and sarrao}.  In this method we measure the mechanical resonances of a sample held lightly between two piezoelectric transducers, and  deduce  the complete elastic stiffness tensor from the spectrum of resonant frequencies by using a computer fitting algorithm. In our RUS apparatus,  the sample and the transducers are mounted inside a home-built measurement cell, which in turn fits inside a Quantum Design Physical Properties Measurement System (PPMS).  We excited the sample using a PZT transducer (Valpey Fisher)  and a Stanford Research Systems (SRS) DS345 function generator and then measured the sample response using a PZT transducer and an SRS SR844 lock-in amplifier.  A Cernox thermometer (LakeShore Cryotronics), located immediately adjacent to the sample, gave us  the sample temperature.  Custom-written LabView software controled the function generator and the lock-in and also instructed the PPMS firmware to measure the temperature and magnetic field. For our RUS measurements, frequency scans (200-1000 kHz) occurred at  approximately 60 temperatures in the range 300-20~K. Before starting each scan, we allowed the sample temperature to stabilize at the pre-programmed value, and we did all measurements in an approximately 1 torr He exchange gas to ensure rapid thermalization between the sample, transducers, and surroundings.

To check for evidence of SnTe piezoelectricity, we performed a modified RUS experiment. We did these experiments exactly as discussed above, except instead of using a PZT transducer as the pick-up transducer, we used a SnTe specimen.  Specifically, with silver epoxy we attached the center conductor of a mini-coax cable to one side of a SnTe specimen and the shield of the coax to the other side.  Then, we connected the other end of the coax to the input
of the lock-in amplifier, exactly as we did for the PZT pick-up transducer.  Finally, we placed the SnTe crystal between the two PZT transducers.  One of the PZT transducers excites the sample, while the other holds it in place.  Performed this way, our RUS uses the SnTe samples as pick-up transducers to self-detect their own resonances.

We also made thermal expansivity/striction, specific heat, magnetic susceptibility, and Hall measurements using the PPMS. We measured the specific heat by a thermal-relaxation technique~\cite{xxx}, and we used a vibrating sample magnetometer (VSM) with a frequency of 20 Hz and a large bore coilset to measure the magnetic susceptibility in fields up to 9~T.  To measure the linear thermal expansion of SnTe and Sn$_{0.995}$Cr$_{0.005}$Te single crystals, we used a capacitive dilatometer\cite{dilatometer} that fits inside the PPMS.  All expansion measurements were done in He exchange gas as the sample slowly warmed from 4~K to 300~K at 0.2~K~min$^{-1}$.  The capacitance was measured using a three terminal capacitance bridge with a resolution of $10^{-7}$ pF, which corresponds to a  sensitivity of  approximately 0.003~\AA{}.

The material also lends itself to analysis with the M\"ossbauer effect, because of the $^{119}$Sn in SnTe. M\"ossbauer studies of the temperature and pressure dependencies can provide site specific information on the isomer shift (valence), the magnetization, the electric field gradient, lattice dynamics, phase transitions, local structure, and mixed composition. M\"ossbauer spectroscopy was carried out with $^{119}$SnTe and $^{119}$Sn$_{0.995}$Cr$_{0.005}$Te samples in the 65--300K range using a Ca$^{119m}$SnO$_3$ source.  A top-loading cryostat was used in which both source and absorbers were kept at the same temperature. A separate measurement in which a standard $^{119}$Sn absorber was kept at ambient temperature. The source temperature was varied in the 50--300K range. This allowed to extract the effective Debye temperature, $\theta_D=300K$, of Ca$^{119m}$SnO$_3$ needed for the analysis of SnTe lattice dynamics.
Measurements with the ferromagnetic Sn$_{0.995}$Cr$_{0.005}$Te sample showed no measurable magnetic hyperfine interaction at the $^{119}$Sn nucleus to the lowest temperature.

\section{Results and Discussion}

\subsection{Transport, specific heat, and magnetization}

Hall effect measurements indicated that our pure SnTe samples had a $p$-type carrier concentration of 3.4$\times$10$^{19}$ cm$^{-3}$ and our Sn$_{0.995}$Cr$_{0.005}$Te crystals had an $n$-type carrier concentration 5.8$\times$10$^{22}$ cm$^{-3}$. Based on transition temperature vs. carrier concentration data in Ref.~\onlinecite{SugaiI}, we expected the structural transition in pure SnTe to occur at aproximately 100~K. The apparent n-type doping in the Cr- doped material is anomalous, and we will return to it later.

To locate $T_c$ more precisely, we first performed resistivity and specific heat measurements. Our resistivity measurement vs. temperature data showed two unambiguous breaks in slope, one at approximately 82 K and the other at 98 K. While the latter was near the expected critical temperature, the former was unexpected and we have no explanation for it. In general, on heating and cooling over this temperature range we were unable to retrace the resistivity curve. The measurements also depended on how and where we attached our leads. As we cooled to 6 K, the resistivity dropped rapidly to zero, in a manner consistent with the onset of superconductivity. Indeed, there have been previous reports of low temperature, carrier concentration dependent superconductivity in SnTe~\cite{ferromagnetic1,allen69,lewis70,mathur73}.


Below 6K our specific heat measurements were fitted by $C_{p}=\gamma T + \beta T^3$, with values of $\gamma$=0.32mJK$^{-2}$mol$^{-1}$ and the inferred Debye temperature,
$\Theta_D$ = 206K, values consistent with previous reports~\cite{heatcapacity}. Using our measurements of  $\gamma$ and the magnetic susceptibility, we computed the Wilson ratio and found it to be approximately 2, the free electron value.  This result prompted two other types of measurements. One type was ARPES. As discussed at length elsewhere~\cite{ARPES}, both above 98~K and below 82~K these measurements clearly indicated the presence of a Fermi surface that is consistent with \textit{ab initio} band structure calculations.

For our second type of measurement, we first tried ferroelectric switching of SeTe (and Sn$_{0.995}$Cr$_{0.005}$Te). We found that any electric field applied to the samples was short-circuited by their high electric conductivities.  Thus, while this short-circuiting excludes the direct measurement of ferroelectricity as a ferroic property,  we tested whether the weaker condition of piezoelectricity could still apply. Our experimental results clearly show that this is not the case: On structurally transforming the elastic stresses do not generate dipole moments.  A literature search failed to find previous reports of  piezoelectric measurements.

From these measurements we concluded that our non-stoichiometric SnTe is a metal, but we still needed to locate the phase transition. In the vicinity of the expected critical temperature we detected no jump in the specific heat. This result is consistent with one previous report\cite{Iizumi} but inconsistent with another~\cite{Hatta}.   Our ARPES measurements however did indicate that a transition likely occurs but the observed changes were too small to infer the symmetry of the low temperature phase. X-ray diffraction measurements indicated a likely transition around 98 K but again out experimental resolution was inadequate to determine the symmetry of the low temperature phase.
From these additional measurements, we assumed that a transition indeed occurs around 98 K, that it is structural, and that the symmetry of the low temperature phase is rhombohedral.

Sn$_{0.995}$Cr$_{0.005}$Te was found to show a ferromagnetic transition at around room temperature, with a low temperature moment consistent with full magnetization of the Cr. It is surprising to find such a high transition temperature in a lightly doped material; we remark that the compound Cr$_2$Te$_3$ is a ferromagnet with a Curie temperature of 325~K. Although we have not observed phase segregation on even microscopic scales, we believe that this is an important hint which will help to resolve results that follow.

\begin{figure}[t]
    \centering\includegraphics[width=\columnwidth]{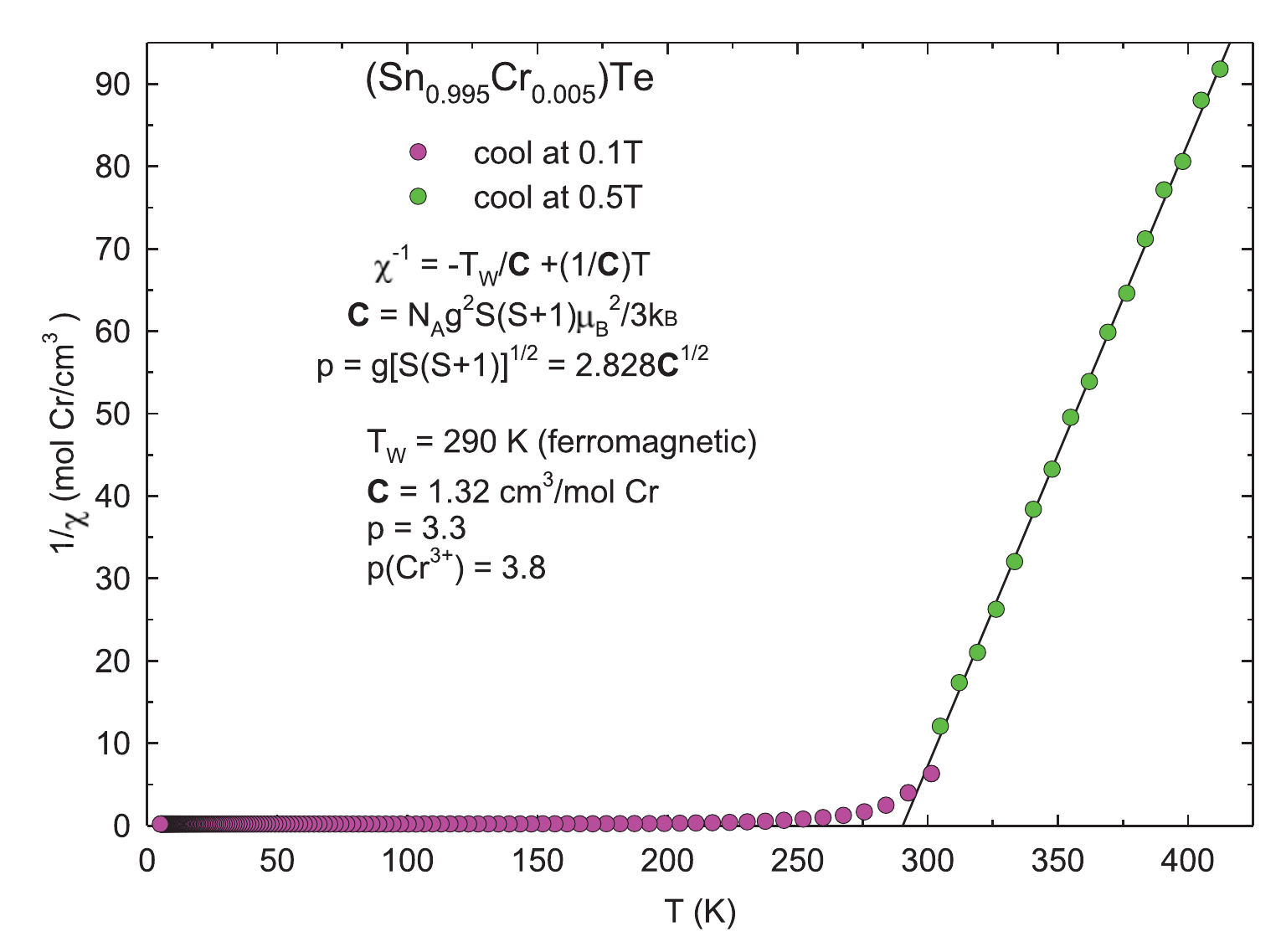}
    \caption{Temperature evolution of magnetization of Sn$_{0.995}$Cr$_{0.005}$Te.
    In this plot, 
    the Curie-Weiss region is only in the temperature range for $T > 300$K.
   The parameters from a fit to the data above this region gives the Weiss constant $T_W = 290$K,
    indicating ferromagnetic ordering. Above the ordering region, the effective number of Bohr magnetons
    is $p = 3.3$. For paramagnetic Cr$^{2+}$ and Cr$^{3+}$, we have $p = 4.9$ and for  $p = 3.8$, respectively.
    Thus $p = 3.3$ is a reasonable value, and indicates the Cr is in the +3 valence state.}
  \label{magnetisation}
\end{figure}

\begin{figure}[t]
    \centering\includegraphics[width=0.8\columnwidth]{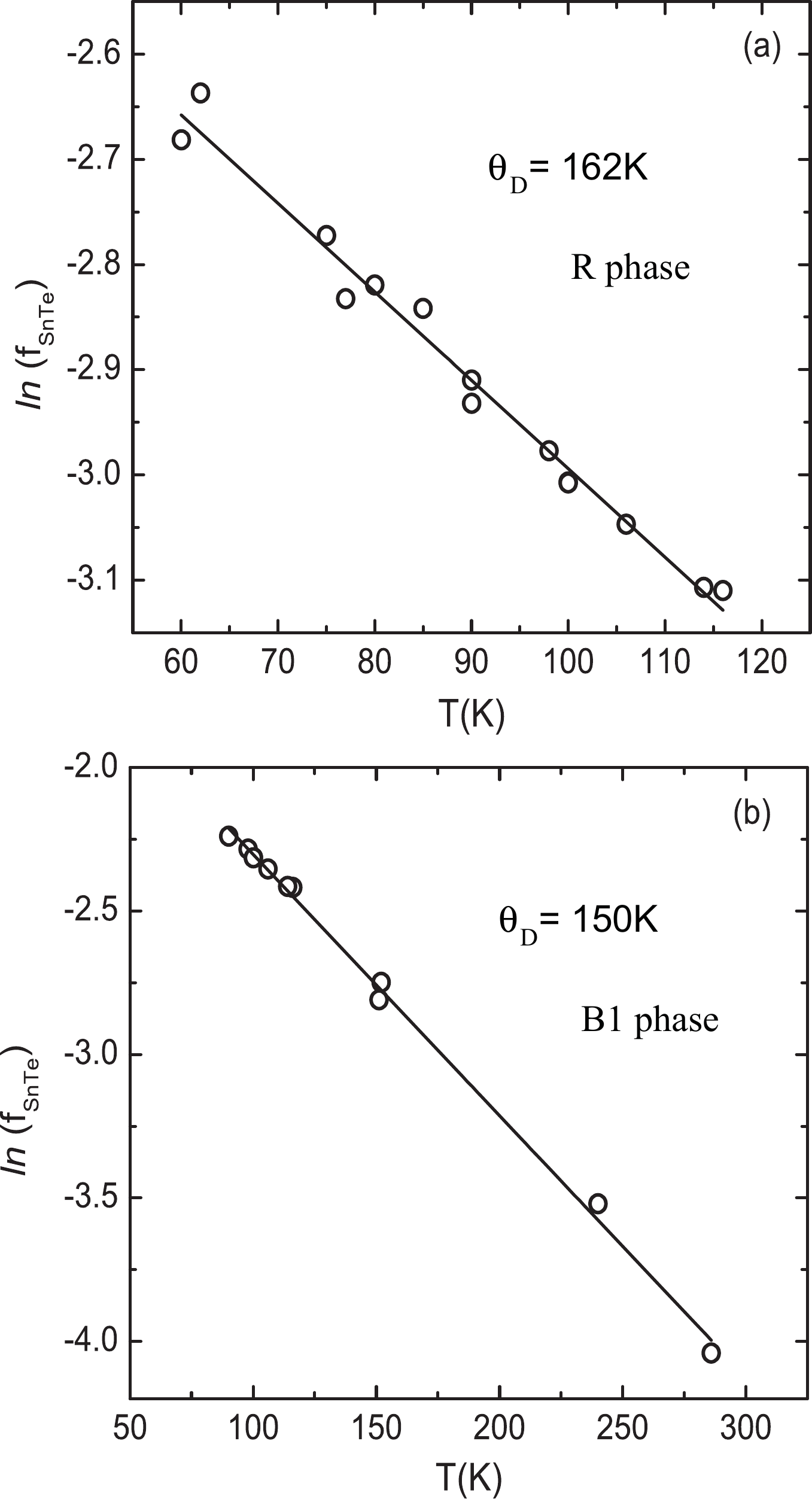}
    \caption{Temperature dependence of the recoilless fraction, $f$, for the temperature regimes: (a) 60--115 K and (b) 90--300K, respectively. The derived $\theta_D$ values., of  162K and 150K, respectively, indicates a hardening of the phonon spectrum in the low-temperature phase.}
  \label{mossbauer}
\end{figure}

\begin{figure}[t]
    \centering\includegraphics[width=0.9\columnwidth]{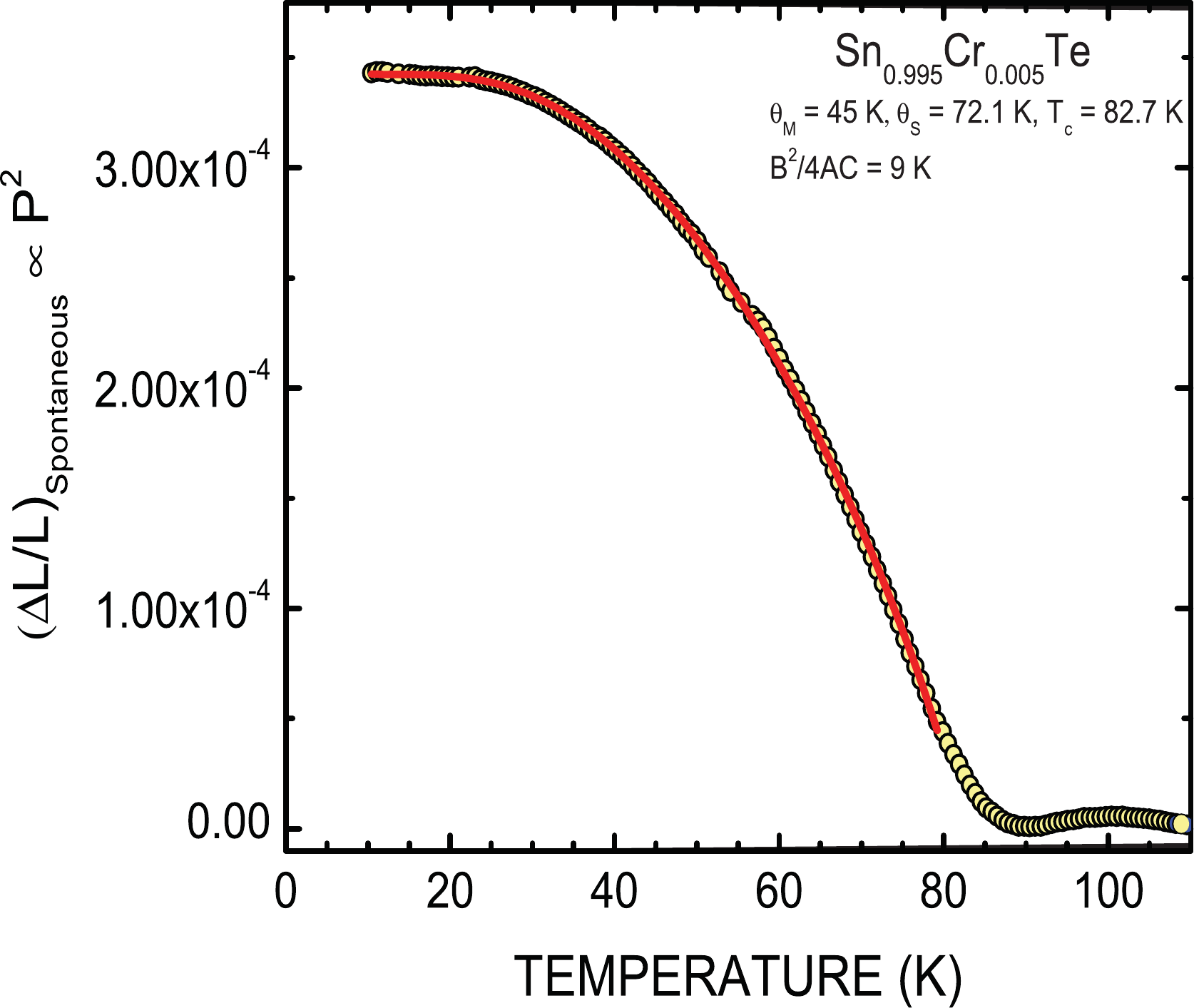}
    \caption{(Color online)  Temperature evolution of the spontaneous strain, $\lbrack\Delta{L}/L\rbrack_\text{spontaneous}$, in the low-temperature R3m phase.  The analysis used to determine the excess strain is discussed in the text.  The curve through the data is the best fit of Eq.~\eqref{spontstrain}. Parameters deduced from this fit are given in the text.}
  \label{excessstrain}
\end{figure}

\subsection{Elastic properties}

We turn now to the elastic properties, which show clear evidence of a structural transition.

Figure~\ref{excessstrain} shows the temperature dependence of the spontaneous strain in the R3m phase of a randomly oriented Sn$_{0.995}$Cr$_{0.005}$Te single crystal.   The spontaneous strain is defined as $\lbrack\Delta{L}/L\rbrack_\text{spontaneous}=\Delta{L}/L-\lbrack\Delta{L}/L\rbrack_B$, where $\Delta{L}/L$ is the measured thermal
strain and $\lbrack\Delta{L}/L\rbrack_B$ is the baseline thermal strain if no phase transition occurred.  We determined $\lbrack\Delta{L}/L\rbrack_B$ by extrapolating the expansion for the high-temperature Fm3m phase into the temperature range of the low-temperature R3m phase.  To do this we assumed that the expansion of the high-temperature phase arises mainly from the anharmonicity of the soft phonon mode that is associated with the phase transition.  We therefore modeled the high-temperature expansion using an Einstein oscillator,
\begin{align} \label{Einstein}
\lbrack{\Delta{L}/{L}}\rbrack_{B} = D + F \coth{(\theta_{M}/T)} \>,
\end{align}
where $\theta_{M}$ is the characteristic temperature of the soft mode and $D$ and $F$ are constants.  We found the value of $\theta_{M}$ using an iterative procedure in which we guessed a value, fitted Eq.~\eqref{Einstein} to the $\Delta{L}/L$ data for the high-temperature phase (in the range $90\leq T \leq 145$~K) to determine $D$ and $F$, and then computed $\lbrack \Delta{L}/L \rbrack_\text{spontaneous}$.  Using different values of $\theta_{M}$, we repeated this procedure until $d\lbrack\Delta{L}/L\rbrack_\text{spontaneous}/dT \rightarrow 0$ at 0~K, as required by the third law of thermodynamics.  In this way, we found  $\theta_{M} =$ 45~K.  This value is in good agreement with the soft-mode frequency measured using Raman spectroscopy~\cite{softmodes}.  As discussed in the theory section, this spontaneous strain is associated with the elastic shear modulus $C_{44}$.

The remaining contribution is the spontaneous strain, which is shown in Fig.~\ref{excessstrain}.
For comparison, in Fig.~\ref{mossbauer} we report results of M\"ossbauer measurements.
In the harmonic approximation using the Debye approximation, the recoilless fraction $f$ can be written as:
\begin{align}
f = \exp \left [
- \frac{6 E_R}{k \theta_D} \left ( \frac{1}{4} + \frac{T}{\theta_D} \int_0^{\theta_D/T} \frac{x \, dx}{e^x-1} \right )
\right ]
\>,
\end{align}
where $E_R$ is the $^{119}$Sn recoil energy (2.57 meV=30.8K/$k_B$). In the high temperature limit, we have
\begin{align}
f = \exp \left (
- \frac{6 E_R}{k} \frac{T}{\theta_D^2}
\right )
\>, \qquad T \ge \frac{1}{2} \theta_D
\>.
\end{align}
The temperature dependent area $A(T)$, of the absorption spectrum is proportional to the recoilless fraction of the sample $f_a(T)$:
\begin{equation}
A(t) = \mathrm{const} \ f_a(T) \>,
\end{equation}
where the constant includes all experimental parameters independent of $T$. The Debye temperature, $\theta_D$, can be extracted from the $T$-linear dependence of $\ln (f_a)$.  In Fig.~\ref{mossbauer}, we show the $T$-dependence of $A$ for two temperature regimes, namely, the 60--115 K and the 90--300K regimes, respectively. Correspondingly, we obtain the $\theta_D$ values as 162K and 150K, respectively, which indicates a hardening of the phonon spectrum in the 60--115 K versus the 100--300 K range.

These results are consistent with the Landau theory discussed in the previous section.
First, we derived a thermodynamic potential $G(P)$ by relaxing the strain under the condition $dG(P,e_{ij})/de_{ij}=0$ where $P$ is determined by the condition of thermal equilibrium $dG(P)/dP=0$ and $e_{ij}$ scales proportional to  $P^{2}$.  We  then used  our experimentally measured thermal expansion for Sn$_{0.995}$Cr$_{0.005}$Te to determine the parameters of the Landau potential.

For the present purposes, we simplify to treat the polarisation and spontaneous strain as scalars, so the free energy becomes
\begin{equation}
\label{free_en}
G (P, e) = \frac{1}{2} \alpha P^2 + \frac{1}{4} \beta P^4 + \frac{1}{6} \gamma P^6 + \delta e P^2 +\frac{1}{2} K e^2 - e \sigma
\end{equation}
Here we choose the model form of Eq.\ref{softmodealpha} for the quadratic coefficient $\alpha$. For convenience, we scale the polarisation to be dimensionless, by setting $\gamma = \alpha (T=0)$. (In the case of a tricritical theory ($\beta = 0$) will fix $P (T=0,e=0) = 1$). $e$ is the dimensionless strain, $\sigma$ the applied stress, and $K$ the relevant elastic modulus. In this case all the material constants $\alpha$, $\beta$, $\gamma$, $\delta$ and $K$ have dimensions of energy per unit volume.

The ground state is determined from the simultaneous minisation of Eq.~\eqref{free_en} with respect to strain $e$ and polarisation $P$. In the absence of applied stress, this leads to a strain $e= \Delta L/L \propto P^2$ which becomes
\begin{equation}
\label{spontstrain}
e= - \frac{b}{2} + \frac{1}{2} \sqrt{ \left (b^2 - 4a \theta_s [ \coth(\theta_{s}/T) - \coth (\theta_{s}/T_{c}) ] \right) } \>,
\end{equation}
where $ b = ( \frac{\delta}{\gamma} ) (\frac{\beta - 2 \delta^2/K}{\gamma})$ and $a = \frac{A \delta^2}{\gamma K}$.
The curve in Fig.~\ref{excessstrain} shows the best fit of \eqref{spontstrain} to the spontaneous strain data for SnTe and Sn$_{0.995}$Cr$_{0.005}$Te.  From this fit we deduced  $a = 1.09\times10^{-8} $~K$^{-1}$ and $b= 6.3\times10^{-4}$,  $\Theta _{s}= 72.1$~K, and $T_{c} = 82.7$~K.   Notice that $b^{2}/4a\approx{9}$~K, indicating that the phase transition is close to tricritical~\cite{Littlewood80,Saljebook}. This analysis allows us to extract the coupling constant to strain, namely $\delta/K = 5.8\times10^{-4}$, and the renormalised quartic coefficient $\beta^\prime/\gamma = (\beta - 2 \delta^2/K)/\gamma = 1.08$.

We checked the validity of these parameters by calculating the temperature evolution of $P = p_{o} \sqrt{e_4}$ where $p_{o}$ is a proportionality factor. The calculated values of $P$ were compared with the experimental ones\cite{Iizumi} and the good agreement affirms that our normalized Landau potential is a good approximation for the excess Gibbs free energy of the transition.

We next estimated an upper bound for the absolute values of the Landau parameters. A direct determination would require experimental data for the excess specific heat of the phase transition in order to calibrate the specific heat jump  $\Delta C_{p}= T \, d^{2}G(P)/dT^{2} |_{T_c}$. Hatta and Kobayashi\cite{Hatta} measured the excess specific heat at temperatures near the transition point, and they found a jump at $T_{c}$ of $0.45\, \text{J mol}^{-1} \text{K}^{-1}$. We were unable to reproduce their data.
Our own resolution was below 0.2Jmol$^{-1}$K$^{-1}$ but we still could not detect a reproducible signal marking the transition in pure SnTe. If we take the value of 0.1Jmol$^{-1}$K$^{-1}$  as an estimate for the excess value at $T_{c}$, we can estimate an upper bound for $A$, specifically,  $A = 0.13\, \text{J mol}^{-1} \text{K}^{-1}$.

\begin{figure}[carrierconcentration]
    \centering \includegraphics[width=\columnwidth]{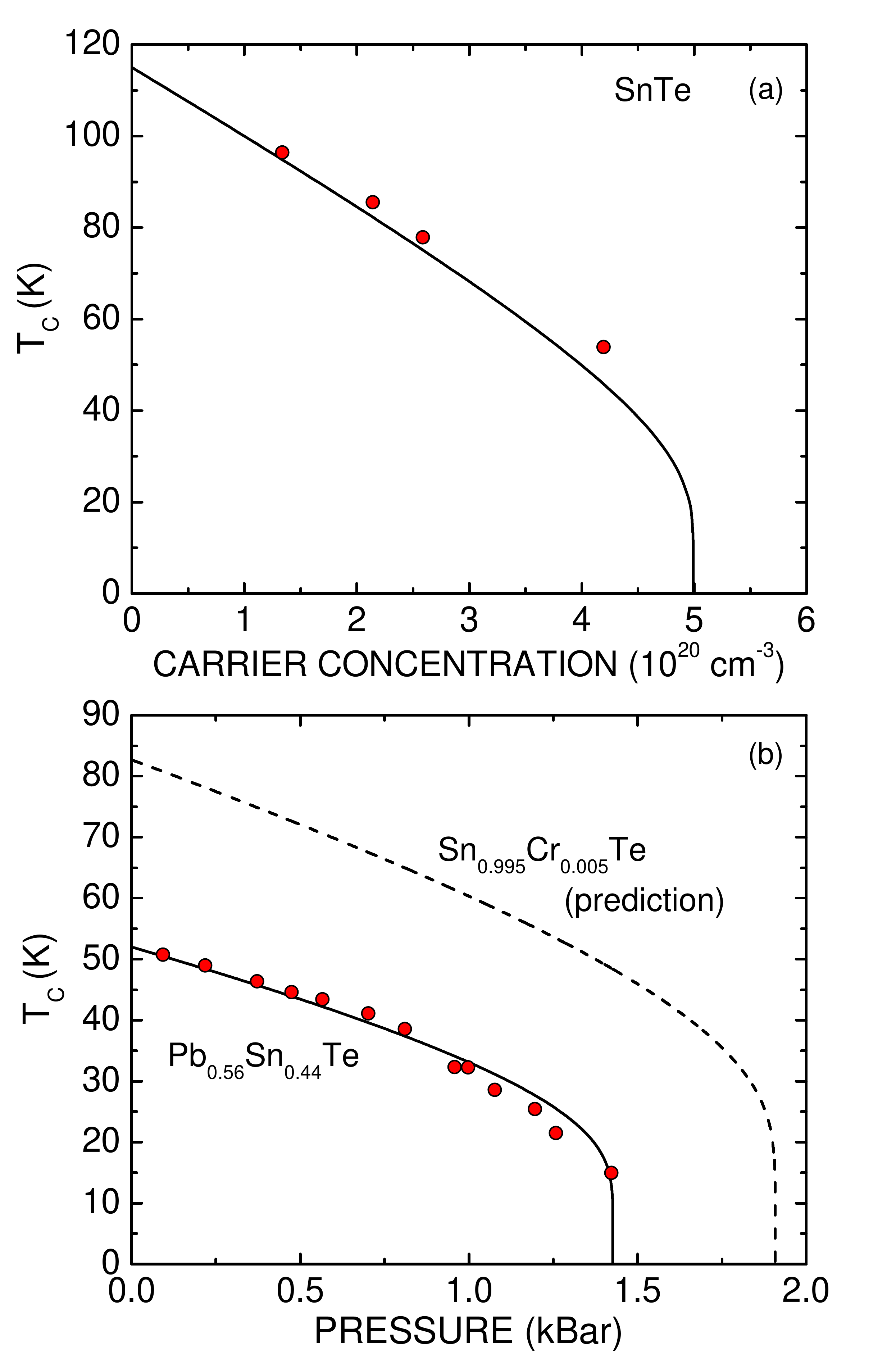}
    \caption{(Color online)  Dependence of the transition temperature on pressure and on the carrier concentration as derived from Landau theory including the quantum saturation (data points from Sugai et al. \cite{SugaiI}). Shown also is the predicted pressure dependence of $T_c$ from the Landau parameters established in this paper.}
    \label{sugaidata}
\end{figure}

It has been known for some time that both increasing carrier concentration and increased pressure suppress the transition, though reliable pressure data has only been obtained on a $Pb_{1-x}Sn_xTe$ alloy.
With the estimated parameters we can now predict the pressure-dependence of the critical temperature. Under an externally imposed stress, there is a Hooke's law strain of $e= \sigma/K$, which contributes a extra positive  quadratic term in the polarisation, and therefore shifts the transition temperature to the temperature where $\alpha + 2 \delta e = \alpha (T) + 2 \delta \sigma/K$ vanishes.
The critical pressure as a function of temperature is then
\begin{equation}
\label{pcrit}
P_c(T) = \frac{A \theta_s}{2(\delta/K)} \left [\coth (\frac{\theta_s}{T_c}) - \coth (\frac{\theta_s}{T_c}) \right]
\end{equation}
With the above parameters, the transition is predicted to vanish beyond a pressure of 1.9$kbar$, shown in Figure\ref{sugaidata}. We are not aware of measurements on pure SnTe, but Sugai et al.\cite{SugaiI} have reported results on the alloy Pb$_{0.56}$Sn$_{0.44}$Te, which has a lower critical temperature and pressure. This data is also shown in Figure\ref{sugaidata}, fit to the same functional form as \eqref{pcrit}, with quite clearly similar parameters to those we extracted for the higher-$T_c$ material.


With a further assumption  we can interpret the carrier concentration dependence from previous measurements, also shown in Fig.~\ref{sugaidata}.
We note that we can analyze the carrier concentration dependence $n$ of the phase transition with the same perturbation approach. The symmetry-allowed lowest-order term for the carrier concentration dependence of $G(P)$ is $kP^{2}n $  ~\cite{saturation} so for a continuous transition the condition for $T_{c}$  becomes
$\alpha ^{*}=\alpha + kn = 0$  or
$T_{c} (n)  = \Theta_{s} /(\coth^{-1}(\coth( \Theta_{s}/ T_{c}(n=0))  - kn))$.
 In Fig.~\ref{sugaidata}, we compare the calculated dependence of the transition temperature on the carrier concentration,  with $k=0.37 \times 10^{20}$ cm$^{-3}$, with experimental data~\cite{SugaiI}.

\begin{figure}[t]
    \centering\includegraphics[width=\columnwidth]{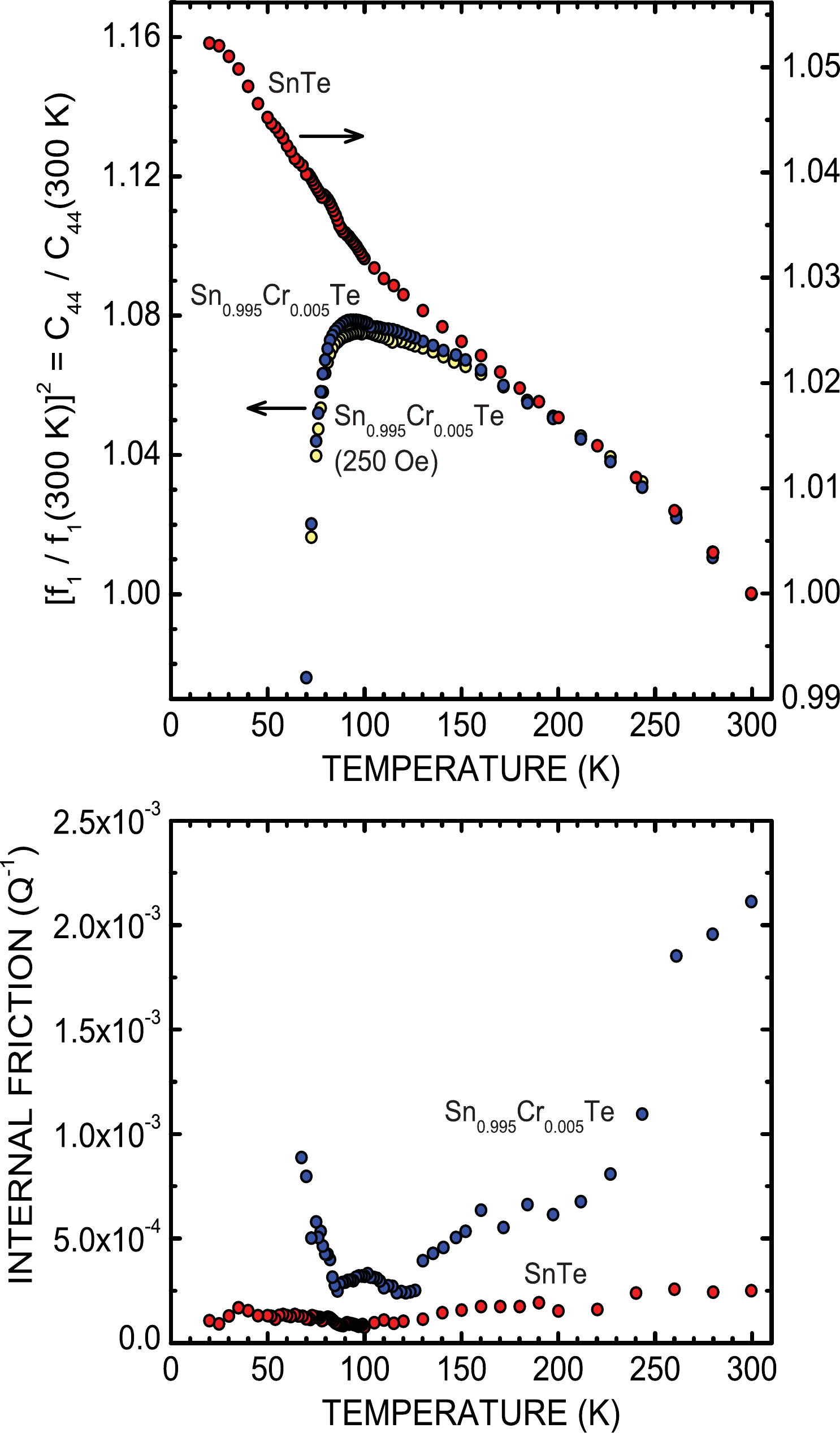}
    \caption{(Color online) Upper panel: Temperature evolution of the shear modulus $C_{44}$, expressed as the square of the lowest resonant frequency of the sample measured in a RUS experiment.  Notice the different scales for the left- and right- ordinate axes. Lower panel: Internal friction (loss) of the modes.}
    \label{shearmodulus}
\end{figure}

A change in the elastic properties is indicated near the transition, and we now turn to direct measurement of the shear modulus from resonant ultrasound.
Figure~\ref{shearmodulus} shows the temperature dependence of [$f_1$/$f_1$(300~K)]$^{2}$, where $f_1$ is the lowest mechanical resonant frequency of the sample measured using RUS, and $f_1$(300~K) is its value at 300~K.  Data are shown for pure SnTe, measured at zero applied magnetic field, and for Sn$_{0.995}$Cr$_{0.005}$Te, measured at 0 and 250~Oe.  For both pure and Cr--doped SnTe, the change in $f_1$ depends only on the value of $C_{44}$ (for example, $df_{1}/dC_{44} =1$ whereas $df_{1}$/$dC^{\prime}=0$ and $df_{1}$/$dB=0$). Thus Fig.~\ref{shearmodulus} shows the temperature dependence of the normalized shear modulus $C_{44}/C_{44}$ (300~K).  Our results are consistent with Beattie's observation~\cite{Beattie} that pure SnTe has no elastic softening at the transition, but rather a small stiffening in the R3m phase below approximately 100~K. In contrast, we find that for Cr--doped SnTe there is a dramatic softening of $C_{44}$ associated with the transition.  The softening is accompanied by a large increase in the damping of the mode, shown in the lower panel of the figure, absent for the pure compound.

We note that well above the transition temperature (for example, $T\geq175$~K),  $C_{44}$ shows the same temperature scaling in both the pure and the Cr--doped materials, yet softening in the Cr--doped material occurs well above~$T_c$.
	
\begin{figure}[t]
    \centering\includegraphics[width=\columnwidth]{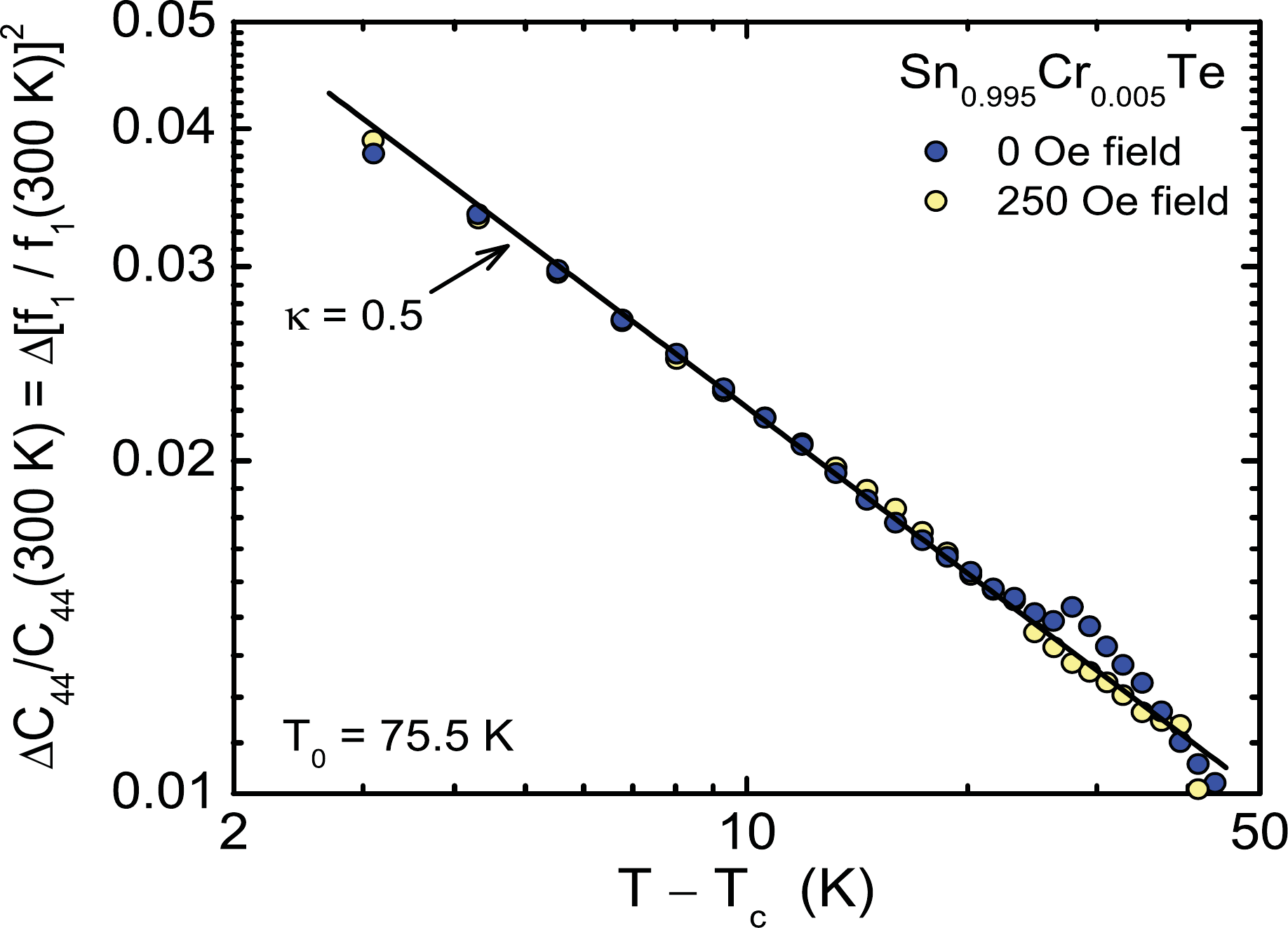}
    \caption{(Color online) Log-log plot of the precursor softening versus the reduced temperature $T - T_{0}$. The theoretical softening is $\Delta C_{44} = A(T-T_{0})^{-\kappa}$ with  $\kappa=0.5$.}
    \label{precursor}
\end{figure}

Note that the softening observed in the Cr-doped material extends from well above the phase transition temperature ($T_c \approx 85 K$) to well below it. Notice further that below $T_c$ the internal friction grows in the Cr-compound and not in the pure material. This makes it clear that this is not the effect of critical fluctuations, but likely a precursor effect due to nucleation and motion of domain walls.
In Fig.~\ref{precursor} we show that the elastic precursor softening $\Delta{C_{44}}$/$C_{44}$ of Sn$_{0.995}$Cr$_{0.005}$Te fits to a power law in reduced temperature $T - T_{0}$, where $T_{0}$ is a temperature some 10 K {\em  below} the critical temperature for the phase transition.  The precursor softening is the difference between the measured temperature dependence of $C_{44}/C_{44}$(300~K) and the baseline $T$--dependence if no phase transition occurred.  For the baseline we used the temperature evolution of $C_{44}/C_{44}$(300~K) in the \emph{high-temperature} phase of pure SnTe, extrapolating these data linearly for temperatures below 110~K so as to exclude the phase transition.

The elastic precursor effect in Cr--doped SeTe is much larger than that usually seen in ferroelastic materials~\cite{precursor}. The temperature dependence of the precursor softening given by theory~\cite{precursor} is
\begin{align}\label{precursoreqn}
\Delta C_{44}/C_{44}\propto{[ (T- T_{0}) /T_{0}]^{-\kappa}},
 \end{align}
where the value of $\kappa$ depends of the dimensionality of the phonon softening.  For the high-temperature phase of Sn$_{0.995}$Cr$_{0.005}$Te, with cubic space group Pm3m, we would expect a value of $\kappa \approx 0.5$, consistent with 3-dimensional softening of the phonons. This is not a critical phenomenon, but represents a three-dimensional softening of the acoustically coupled soft mode.

\section{Concluding Remarks}

The transport and (lack of piezoelectricity) measurements reported here and the recently reported ARPES measurements are sufficient to establish that SnTe is a metal and not a semiconductor. Hence, it is not a ferroelectric as commonly assumed. Without doubt the observed metallic behavior, instead of the semiconducting behavior predicted by band structure calculation, is a consequence of the non-stoichiometry. We see this result as significant because we are unaware of any published claim of having prepared this material stoichiometrically.

We note a number of our measurements produced inconclusive results about the existence, symmetry, and location of the phase transition. We believe this is partially a consequence of the subtlety of the transition. Although we have no direct evidence,  we  also believe some measurements, in particular, those of the low temperature elastic constants, were influenced by inhomogeneities in the material. Likely, SnTe is not mono-domained. In this regard, we remark that our ultrasound measurements confirmed Beattie's observations\cite{Beattie} that low-temperature SnTe does not soften elastically but instead \emph{hardens}. The absence of softening indicates that mobile twin boundaries do not nucleate at the transition point.  Such behavior is extremely unusual, having apparently so far been observed only in martensitic Cu$_{74.08}$Al$_{23.13}$Be$_{2.79}$ where the immobility of the twin boundaries is caused by strong dislocation pinning~\cite{SaljeII}. All other martensitic transformations studied to date show strong elastic softening near $T_c$, similar to what we observed in Sn$_{0.995}$Cr$_{0.005}$Te (Fig.~\ref{shearmodulus}).

Several physical scenarios can explain the observed hardening in SnTe.  One is that the twin boundaries are heavily pinned by dislocations as in the case of Cu$_{74.08}$Al$_{23.13}$Be$_{2.79}$~\cite{CarpenterI}.  In the present case, however, this scenario is unlikely because by symmetry the phase transition in SeTe does not generate dislocations, and other pinning mechanisms for domain walls, such as Peierls pinning, are too weak to prevent the movement of ferroelastic domain boundaries.  A second more probable scenario is the absence of domain boundaries. This scenario implies what we measure at the transition is the intrinsic hardening of the shear modulus.  In contrast to SnTe, we note that Cr--doped SnTe near $T_c$ exhibits the strong elastic softening which we suggest is related to the nucleation of ferroelastic domain boundaries and to their mobility under the externally applied stress of the ultrasound~\cite{ferroelastic}.

We noted earlier that our Cr-doped samples show bulk ferromagnetism at a temperature close to that expected for the compound Cr$_2$Te$_3$. We have seen no microstructural evidence for phase segregation, but it is likely that a nano-scale second phase plays an important role in nucleating domain boundaries and enhancing their mobility. Cr$_2$Te$_3$ is a good metal, and metal inclusions in the semimetallic SnTe would also produce anomalous magnetotransport. For practical purposes, however the Cr-SnTe samples show homogeneous ferromagnetism and ferroelasticity. Coupling between the magnetic and structural order is nonetheless weak, and probably indirect.

In closing, we comment that while future experiments may clarify the material science of the elastic properties, it is interesting to propose the more fundamental question, ``If stoichiometric SnTe existed, would it be ferroelectric at low temperatures?'' If so, then some amount of carrier doping, perhaps, an infinitesimal amount, would generate a very unusual metal-insulator transition, that is, one accompanied by the creation or destruction of electrical polarization.

\acknowledgements{This work was supported in part by the Department of Energy's Laboratory Directed Research and Development Program.}

\bibliography{snte 07}

\end{document}